# The Essential Work of Fracture Parameters for 3D printed polymer sheets

I.I. Cuesta [(1)], E. Martinez-Pañeda [(2)], A. Díaz [(1)], J.M. Alegre [(1)]

[(1)] Structural Integrity Group, Universidad de Burgos, Avda. Cantabria s/n, 09006 Burgos. SPAIN
[(2)] University of Cambridge, Department of Engineering, Trumpington Street, Cambridge CB2 1PZ, UNITED KINGDOM

Telephone: +34 947 258922; e-mail: iicuesta@ubu.es

**Abstract.** Additive manufacturing is becoming increasingly popular in academia and industry. Accordingly, there has been a growing interest in characterizing 3D printed samples to determine their structural integrity behaviour. We employ the Essential Work of Fracture (EWF) to investigate the mechanical response of polymer sheets obtained through additive manufacturing. Our goal is twofold; first, we aim at gaining insight into the role of fibre reinforcement on the fracture resistance of additively manufactured polymer sheets. Deeply double-edge notched tensile (DDEN-T) tests are conducted on four different polymers: Onyx, a crystalline, nylon-reinforced polymer, and three standard polymers used in additive manufacturing – PLA, PP and ABS. Results show that fibre-reinforcement translates into a notable increase in fracture resistance, with the fracture energy of Onyx being an order of magnitude higher than that reported for non-reinforced polymers. On the other hand, we propose the use of a miniature test specimen, the deeply double-edge notched small punch specimens (DDEN-SP), to characterize the mechanical response using a limited amount of material. The results obtained exhibit good alignment with the DDEN-T data, suggesting the suitability of the DDEN-SP test for measuring fracture properties of additively manufactured polymers in a cost-effective manner.

## 1. Introduction

Additive manufacturing is an emerging technology that is becoming an alternative method for the manufacture of components in a wide range of industries. One of the main advantages of additive manufacturing is the reduced time elapsed from the conception of the component to its final manufacture, as it does not require the design and production of special tools through other production processes. The freedom allowed in the design stage is maximal, to the extent that development engineers can create geometries that cannot be obtained by other methods in a cost-effective manner. Consequently, one of the advantages of additive manufacturing is the high profitability in the production of short series or prototypes, especially when dealing with complex geometries. Numerous 3D printing technologies have been proposed to date, each of them with their advantages and disadvantages. Five techniques appear to enjoy greater popularity: fused deposition modelling (FDM), laser deposition modelling (LDM), selective laser sintering (SLS), stereolithography (SLA) and multi-jet printing (MJP). Due to its low upfront cost, fused deposition modelling (FDM) is one of the techniques with greater future prospects. This technology is based on heating a polymer above its glass transition temperature, and depositing it layer by layer with a nozzle. The main drawbacks of FDM are the notable surface roughness levels and, common to other additive manufacturing techniques, the poorer mechanical performance of the samples (relative to traditional manufacturing methods). Not surprisingly, a vast amount of literature has been devoted to the characterization of the mechanical properties of specimens manufactured by FDM. Most works aim at assessing the role of printing process parameters on the mechanical properties (see, [1-10], and references therein) but there are also studies related to fracture and fatigue properties [11-13], and high temperature performance [14-15].

The present work has a twofold objective. First, we aim at gaining quantitative insight into the role of fibre-reinforcement on the enhancement of the fracture resistance of additively manufactured polymer sheets. To this end, we make use of the Essential Work of Fracture (EWF), see Section 2.2, to obtain the fracture and plastic properties of different types of polymers. Specifically, we take as benchmark material a nylon-based crystalline polymer reinforced with short carbon fibres named Onyx. The mechanical and fracture properties of Onyx are compared to those obtained for other typical 3D printing polymers, both crystalline and amorphous. These are polylactic acid (PLA), polypropylene (PP), and acrylonitrile butadiene styrene (ABS). The second objective of the paper aims at assessing the suitability of the so-called Small Punch Test (SPT) to characterize the mechanical behaviour of additively manufactured polymers. The SPT has proven to be a reliable experiment for characterizing the mechanical and fracture properties of metallic materials (see, e.g., [16-17]), and its use has been recently extended to polymer testing [18]. The test employs very small specimens, making it very cost-effective and particularly suitable for situations where a limited amount of material is available. We investigate the viability of the method by comparing with the results obtained from the standardised deeply double-edge notched tensile (DDEN-T) experiment. The aim is to elucidate the role of two potential obstacles. The first one involves the specimen-to-specimen variability, an effect that is already present in additively manufactured samples (more defects and heterogeneities) and that can be magnified further with the use of miniature specimens. The second issue in extrapolating to standardised experiments lies in the different stress triaxiality conditions and loading configuration, which can bring changes in the damage mechanisms at play. We aim at gaining insight into these effects, and discuss their influence on the viability of the approach proposed.

## 2. Methodology

We subsequently proceed to describe the materials employed in the study (Section 2.1), the Essential Work of Fracture method employed (Section 2.2) and the different experimental tests adopted (Section 2.3).

### 2.1 Materials and analysis of porosity

The influence of fibre-reinforcement will be assessed by conducting experiments on a chopped carbon fibre reinforced nylon with commercial name Onyx. With a flexural strength of 81 MPa, this material is roughly 1.4 times stronger and stiffer than ABS, and can be reinforced with any continuous fibre. Onyx is supplied as a continuous filament ready to be printed using FDM. Figure 1 shows the aspect of the short fibres embedded in the polymer matrix, having an approximate length of 61.7 microns and a diameter of approximately 8.2 microns. The fibres provide an additional stiffening, which makes Onyx 3.5 stiffer than regular nylon. Onyx has a density of 1.2 g/$cm^3$, a Young's modulus of 1.4 GPa, and an elongation at failure of 58%.

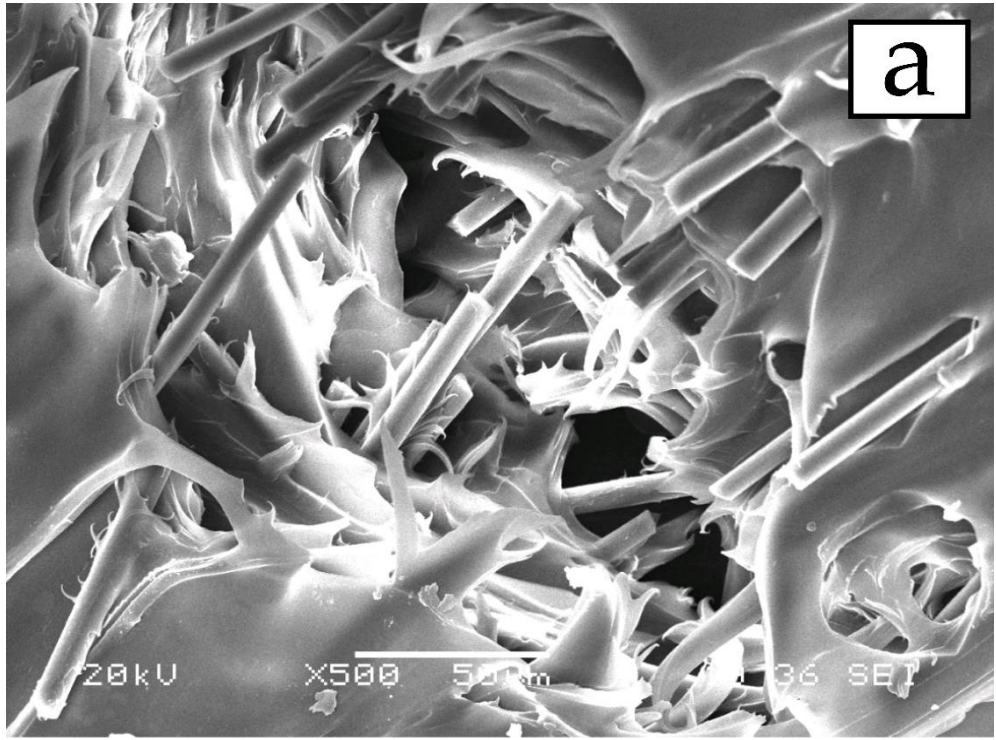

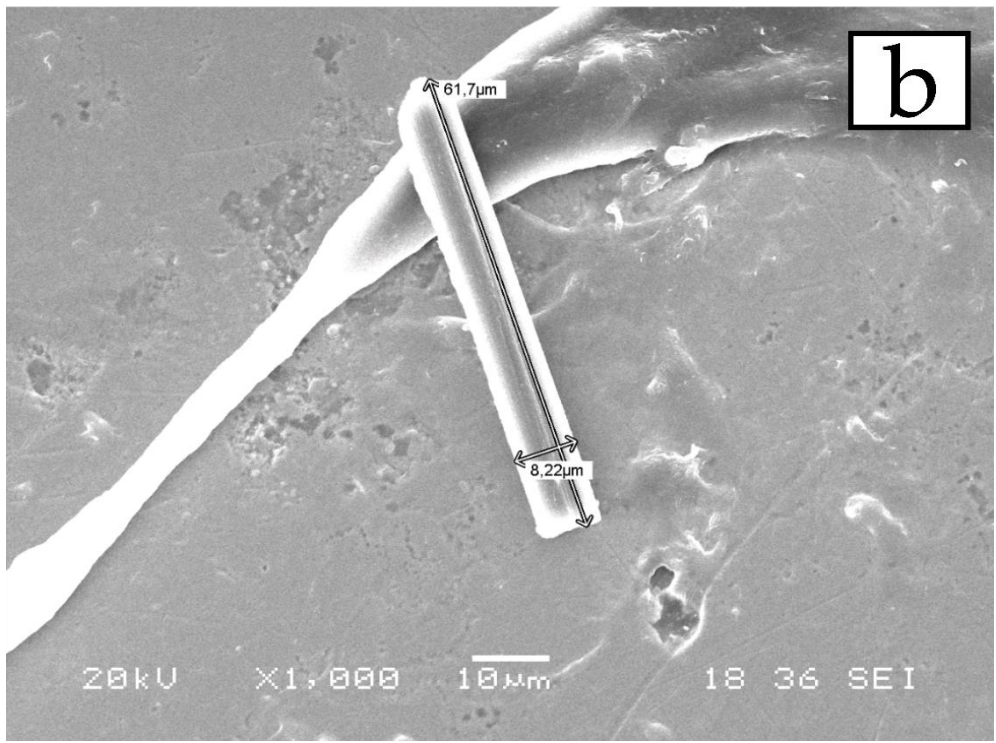

*Figure 1. Aspect of the short fibres embedded in the nylon matrix: (a) general view, and (b) detailed view.*

We use computerized axial tomography to determine the degree of initial porosity of the FDM 3D printed samples. The details of the fracture process zone in a deeply double-edge notched tensile (DDEN-T) specimen are shown in Figure 2. The deposition of the material during the printing process can be clearly observed. As expected, the different passes of the extruder leave small empty gaps, reducing the density of the final specimen.

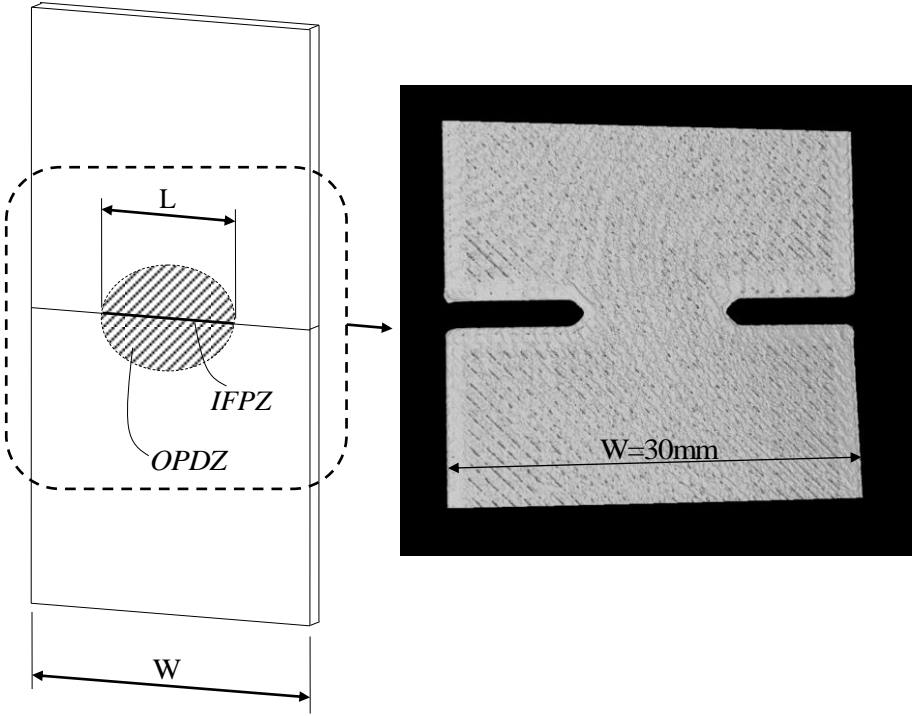

*Figure 2. Deeply double edge notched tensile specimen. The outer process dissipation zone (OPDZ) and the inner fracture process zone (IFPZ) are marked.*

We repeat the same procedure for the experimental setup proposed to characterize additively manufactured polymers, the deeply double-edge notched small punch (DDEN-SP) test. As described in Section 2.3, the DDEN-SP experiment is an extension of the standard Small Punch Test (SPT) for metals, with two deep side notches and a shape that resembles the standardised DDEN-T specimen. The computerized axial tomography image of the DDENT-SP is shown in Figure 3. The pattern of deposition of the material (45º in two directions) and the final porosity obtained are similar to those found for conventional test specimens (DDEN-T). Accordingly, the porosity levels are expected to be similar in both DDEN-T and DDEN-SP specimens. This is confirmed by detailed porosity measurements. From 10 porosity measurements, a porosity mean value of 2.2 ± 0.1% has been estimated, wherein the value of 100% corresponds to completely dense material.

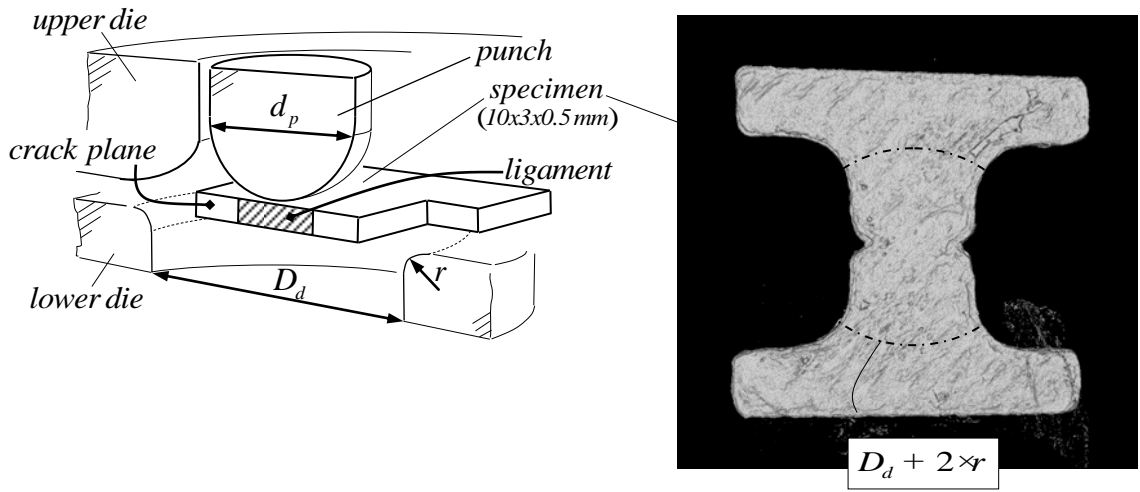

*Figure 3. Small punch test device and DDEN-SP specimen.*

The performance of this nylon-matrix carbon fibre reinforced polymer is compared to three polymeric materials that are commonly used in FDM additive manufacturing: polylactic acid (PLA), polypropylene (PP), and acrylonitrile butadiene styrene (ABS). PLA and PP are crystalline while ABS is amorphous. PLA is a thermoplastic with a density of 1.25 g/$cm^3$, Young's modulus of 3.5 GPa, and elongation at failure of 6%. On the other hand, ABS exhibits a greater variability, with a density that goes between 1.03 and 1.38 g/$cm^3$, a Young's modulus ranging between 1.7 to 2.8 GPa, and an elongation at failure between 3% and 75%. And PP has a density of approximately 0.9 g/$cm^3$, a Young's modulus in the range of 1.1 to 1.6 GPa and an elongation at failure that can go from 100% to 600%. All the materials considered in this study were produced with the same technique.

**2.2 Essential Work of Fracture**

Broberg [19,20] proposed, for the first time, the Essential Work of Fracture (EWF) method to characterize metals and alloys. This technique was later extended to the characterization of polymers by Mai and Cotterell [21,22]. By using DDEN-T specimens, the EWF has been successfully used to quantitatively measure fracture properties in thin polymer sheets; see, e.g., [23,24] and references therein. The EWF method is particularly suited to characterize the fracture behaviour of polymer sheets with thicknesses lower than 2 mm.

The foundation of the EWF method is the division of the energy consumed during the ductile fracture of pre-cracked specimens ($W_f$) into two terms: the essential work ($W_e$) and the plastic work ($W_p$). The former, $W_e$, represents the energy required to create the new fracture surfaces, which can be related to the inner fracture process zone (IFPZ). The latter, $W_p$, is non-essential work, as it comprises the energy employed in general plastic deformation and the dissipation process, depending on the geometry of the deformed region. This plastic work is thus related to the outer process dissipation zone (OPDZ). Both terms are a function of the specimen ligament as expressed in equation (1). Hence, they are usually divided by the ligament section ($L \cdot t$) to use the specific work terms, ($w_f$, $w_e$ and $w_p$), as shown in equation (2),

$$W_f = W_e + W_p = w_e \cdot L \cdot t + \beta \cdot w_p \cdot L^2 \cdot t \tag{1}$$

$$w_f = w_e + \beta \cdot w_p \cdot L \tag{2}$$

where $t$ is the specimen thickness, $L$ is the ligament length and $\beta$ is the shape factor corresponding to the geometry of the outer plastic dissipation zone [25]. The term $w_e$ can be seen as roughly equivalent to the fracture toughness. For a correct use of the EWF method, self-similarity between load-displacement curves must be achieved [25,26].

For polymer films, the advantage of the EWF method using DDEN-T specimens compared to the J-Integral procedure is, in many cases, its experimental simplicity. An inconvenience arises when there is not enough material available to extract the conventional DDEN-T specimens. We shall explore the viability of extending the success of the Small Punch Test (SPT) to the analysis of additively manufactured polymers. In addition, the EWF and conventional DDEN-T samples will be used to quantify the improvement in fracture properties gained by the use of fibre reinforcement.

**2.3 Experimental tests**

The standard testing example to measure the EWF parameters in polymers is the deeply double-edge notched tensile (DDEN-T) specimen shown in Figure 2. The test is conducted in agreement with the usual procedures (see, e.g., [26] for details) and no special measures are taken. The experiments are performed using a MTS Criterion 43 electromechanical Universal Test System machine, with 10 kN of load capacity. All samples are printed with a width ($W$) of 30 mm.

We propose, as an alternative to the DDEN-T specimen, the use of the deeply double-edge notched small punch (DDEN-SP) samples – see Figure 3. The ability of the Small Punch Test (SPT) in measuring the EWF parameters is examined by modifying the standard SPT specimen so as to mimic the conditions of the DDEN-T experiment [27]. The main enhancement is the introduction of two side notches. Although the small punch test has been used in numerous studies with the objective of mechanically characterizing a material sample of small dimensions [28-30], few of these studies have considered pre-notched SPT specimens [31,32]. The tests are conducted in the configuration outlined in the left side of Figure 3, involving a high-strength punch of diameter $d_p = 2.5\,mm$ and a lower die with a hole with diameter $D_d = 4\,mm$ and fillet radius $r = 0.5\,mm$. Unlike the DDEN-T samples, the DDEN-SP specimens are manufactured with a width of 3 mm. In both DDEN-T and DDEN-SP cases, the samples were additively manufactured directly on their final form. A thickness of 0.5 mm was achieved directly from the FDM manufacturing for all the DDEN-SP specimens, avoiding the need for any polishing or machining process. The remainder details of the testing configuration follow those described in the CEN code of practice for small punch testing of metallic materials [33]. It is noted that reproducibility of the DDEN-SP tests requires a good degree of precision in the FDM process.

Common to both DDEN-T and DDEN-SP specimens, the notches are sharpened by means of a razor blade. As elaborated in Section 3.1, different notch lengths are considered. Given that the specimen width is reduced by different notch sizes, the length of the ligament must vary in such a way that plane stress conditions are still verified. Thus, the range employed in the EWF method is limited by a maximum ligament of $L \geq 3 \cdot t$ and a minimum of $L < W/3$ [25]. All DDEN-T and DDEN-SP tests were performed at room temperature with a test rate of 0.5 mm/min.

## 3. Results and discussion

We shall first address the results related to the analysis of the influence of the fibre reinforcement on the EWF parameters of additively manufactured polymers (Section 3.1). Afterwards, we assess the capabilities of the DDEN-SP testing procedure in evaluating the mechanical and fracture properties of polymeric materials obtained by additive manufacturing (Section 3.2).

### 3.1 Analysis of the fibre-reinforcement effect

Consider first the standardised DDEN-T experiment. We compare the fracture performance of Onyx with the non-reinforced materials: ABS, PP and PLA. Figure 4 shows the results obtained for different notch lengths. As evident from Figure 4, the self-similarity condition is fulfilled; in other words, a similar shape of the force-displacement curve is measured for the different notch lengths. Regarding the magnitude of the maximum load, Onyx and PLA exhibit similar values, which are roughly 1.5 times higher than those that are attained with PP and PLA. On the other hand, the displacement at final failure is of the same order for Onyx, ABS and PLA, but is one order of magnitude higher for PP. The higher ductility of PP is significant and consistent with its higher elongation at break (see Section 2.1). We follow Ref. [34] and quantify the ductility by defining a ductility level $D_L$ as a function of the displacement at rupture $d_r$ and the ligament length $L$ as

$$D_L = d_r / L \tag{3}$$

The ductility level is used to characterize the different fracture behaviours observed in polymer fracture. Thus, $D_L < 0.1$ indicates brittle fracture, $0.1 < D_L < 0.15$ is the regime of ductile instability, the range $0.15 < D_L < 1$ is known as post-yielding, blunting is characteristic of $1 < D_L < 1.5$, and necking is the main dominant failure mechanism when $D_L > 1.5$. The value of $D_L$ is obtained for each test, and found to be within 0.1-0.15 (i.e., ductile instability) for ABS, PLA and Onyx. However, in the case of PP, the ductility level was found to lie between 1 and 1.5, indicative of failure accompanied by extensive plastic deformation at the crack tip, without steady crack propagation.

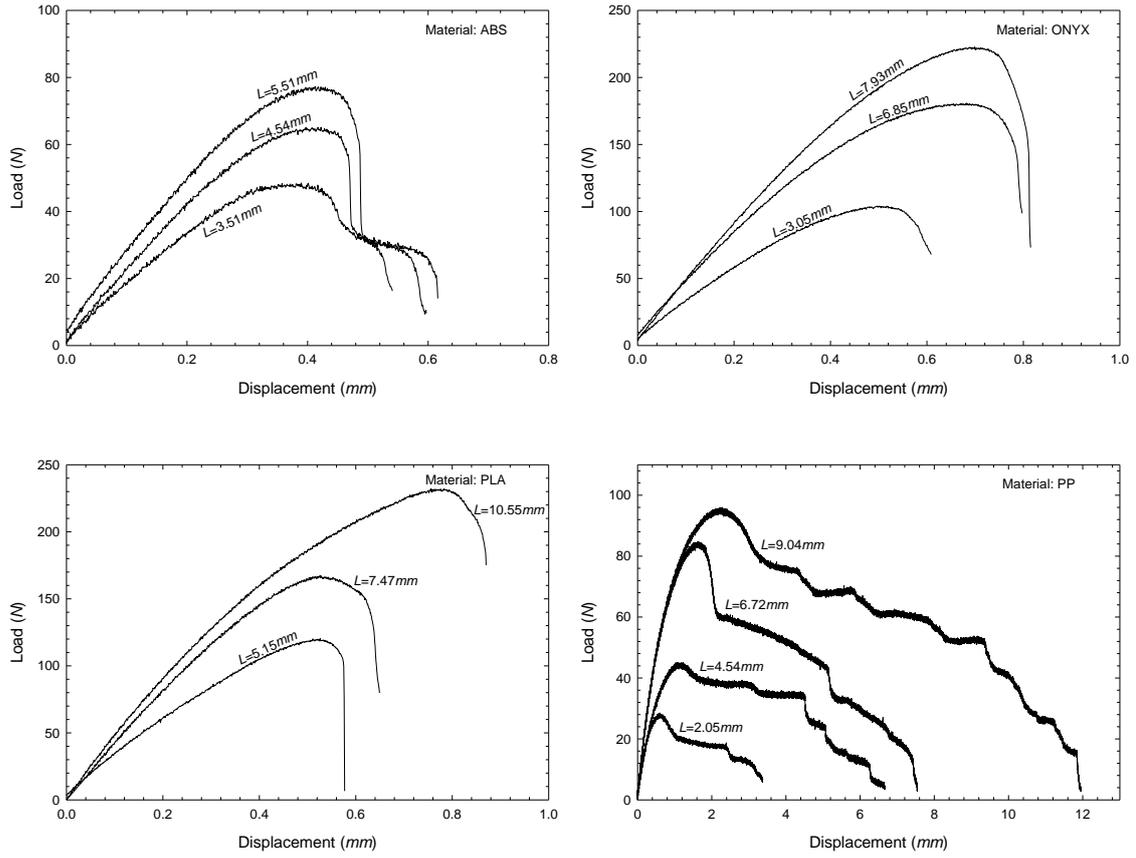

*Figure 4. Load-displacement curves obtained from the DDEN-T specimens for, respectively, ABS, Onyx, PLA and PP.*

We proceed to conduct the EWF analysis for each material. Figure 5 shows the results of the specific work of fracture for the four materials analysed. The figure includes the linear regression of the data, as well as the 95% confidence and prediction bands, according to equation (2). The values obtained for $\beta w_p$, $w_e$ and $R^2$ are listed in Table 1.

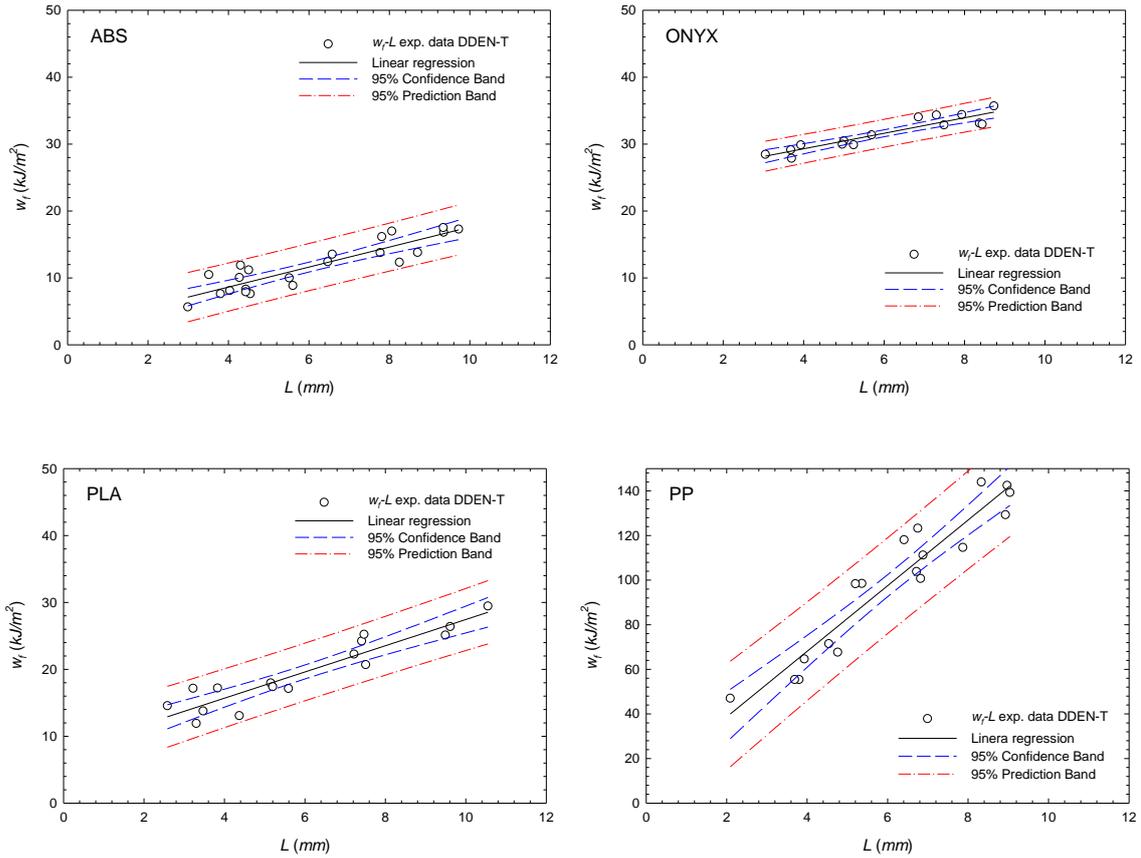

*Figure 5. Specific work of fracture obtained from the DDEN-T specimens for, respectively, ABS, Onyx, PLA and PP. The linear regression of the data, along with the 95% confidence and prediction bands are also shown.*

First, note that, consistent with the ductility analysis above, the plastic term $\beta w_p$ is substantially larger for PP than for the other materials. Very little differences between materials are observed in the term $R^2$, which is in all cases close to 1. The degree of reproducibility is therefore satisfactory. More interestingly, the term $w_e$, which can be assimilated as the initiation toughness, shows important differences between the fibre-reinforced material, Onyx, and the rest. Specifically, the toughness of Onyx is 24.14 kJ/ m$^2$, about an order of magnitude higher than ABS. Thus, the present analysis not only provides with the estimation of the mechanical and fracture properties of Onyx but also quantifies the fracture resistance enhancement provided by fibre-reinforcement, relative to other widely used additively manufactured polymers.

*Table 1. Essential Work of Fracture Parameters.*

| Material | $\beta \cdot w_p$ [$kJ/m^3$] | $w_e$ [$kJ/m^2$] | $R^2$ |
|----------|---------------------|-----------------|-------|
| ABS      | 1.49                | 2.66            | 0.81  |
| ONYX     | 1.14                | 24.14           | 0.87  |
| PLA      | 1.96                | 7.84            | 0.87  |
| PP       | 14.71               | 9.23            | 0.91  |

## 3.2 Assessment of the suitability of DDEN-SP testing

Let us address now the suitability of the miniature DDEN-SP experiment for the characterization of additively manufactured polymer sheets. This proof-of-concept exercise will be conducted with the fibre-reinforced Onyx material. The force versus displacement curves obtained with different notch lengths are shown in Figure 6. It can be clearly seen that the similarity condition between curves for different ligament sizes is fulfilled, as in the standardised DDEN-T experiments. In agreement with expectations, and with the DDEN-T results, the maximum load level increases with the ligament size. Comparing the shape of the force-displacement curves for small punch specimens to those for DDEN-T, a different evolution is found. This finding is attributed to the fracture process at the ligament, as the DDEN-SP specimens do not have the same behaviour as the DDEN-T specimens when the ligament length approaches zero. In a DDEN-T sample, if $L \rightarrow 0$, the load-displacement curve approaches zero ($W_f \rightarrow 0$), because the specimen is separated into two halves, and the force during the test is zero. However, in a DDEN-SP specimen, if $L \rightarrow 0$, the load-displacement curve does not approach zero, because the punch must plastically deform the two halves of the sample to pass through them to complete the test. We make use of the ductility index defined in Equation (3) to assess the role of the stress triaxiality and loading conditions inherent to the DDEN-SP experiments. An average ductility level $D$ can be estimated from the results shown in Figure 6, and compared with the one obtained from the Onyx measurements with DDEN-T samples. As listed in Table 2, the ductility level attained in DDEN-SP is more than four times higher than that attained with DDEN-T. According to the classification outlined in above, the DDEN-SP specimens fail in the post-yielding regime, while the standardised DDEN-T samples lie at the border between the ductile instability and the post-yielding regimes. Thus, differences exist in the ductility levels, which do not compromise the use of the EWF methodology, but will hinder a quantitative correlation between the values of plastic work measured in the two tests, as discussed below.

*Table 2. Average ductility level in the DDEN-T and DDEN-SP tests as measured for Onyx.*

| *Onyx* | DDEN-T | DDEN-SP |
|---|---|---|
| Ductility level, *D* | 0.14 | 0.65 |

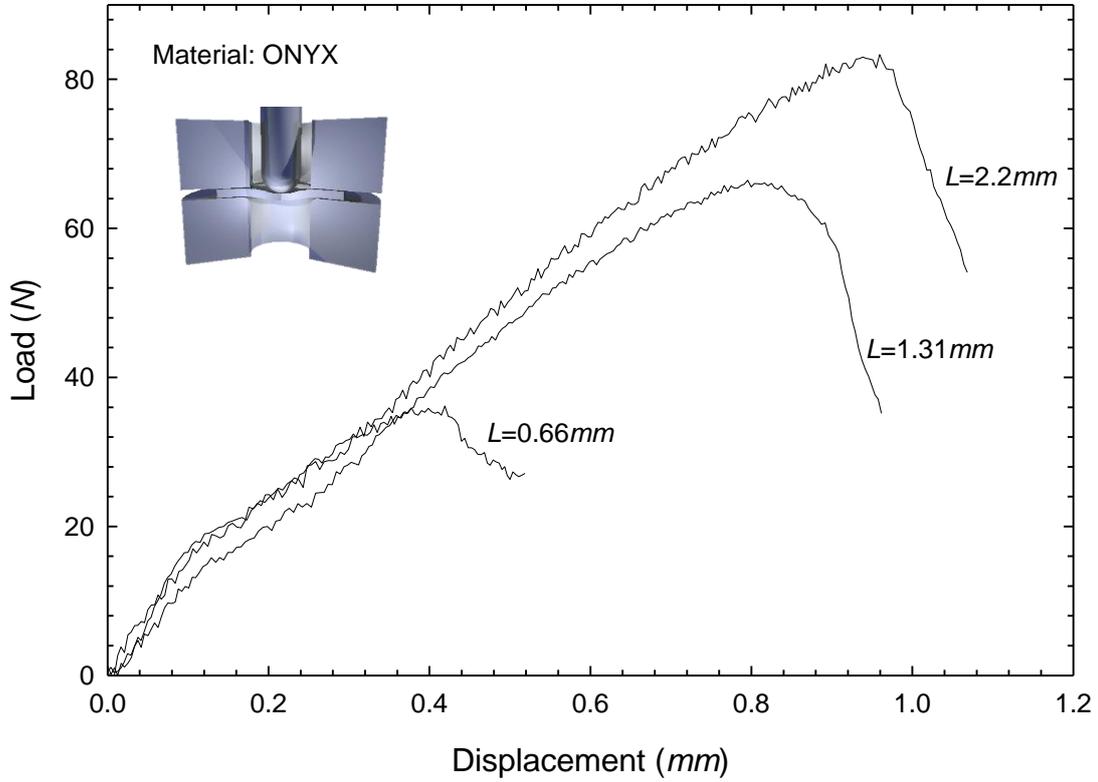

*Figure 6. Load-displacement curves obtained from DDEN-SP specimens for Onyx.*

Figure 7 shows the final features of the DDEN-SP broken specimens, and the zone corresponding to the fracture process can be identified at a glance. A fracture behaviour similar to that of the DDEN-T specimens was found (crack propagation in mode I), and thus, the upper and lower limits of the $w_f - L$ values can be established, as well as the essential and plastic work. Thus, the DDEN-SP experiment appears to be suited to extract the EWF parameters, enabling the assessment of the mechanical behaviour of Onyx. A comparison with the results from the standardised DDEN-T tests is shown below.

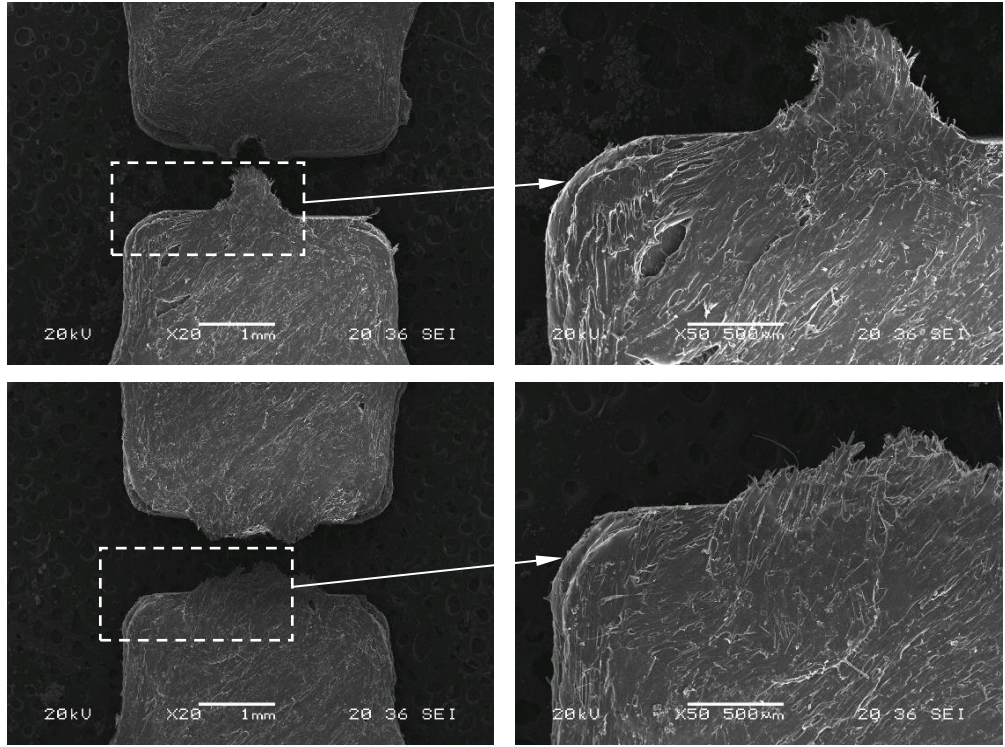

*Figure 7. Detail images of the fracture process in deeply double edge notched small punch specimens for Onyx.*

Figure 8 shows the specific work of fracture, $w_f$, obtained from both DDEN-T and DDEN-SP experiments on Onyx, as a function of the ligament length, *L*. As with Figure 5, the $w_f$ versus *L* results are accompanied by the linear regression of the data, as well as the 95% confidence and prediction bands. In addition, a table is included with the values inferred for the plastic work $\beta w_p$, the initiation toughness $w_e$ and the parameter $R$. We note in passing that, for the DDEN-SP specimens, it is not possible to apply the conventional criteria to establish the upper ($L<W/3$) and lower ($L \geq 3 \cdot t$) limits for the $w_f - L$ values; the DDEN-SP specimens used have a width $W = 3\,mm$, so the limits overlap. To overcome this problem, we choose to make limits consistent with the miniature specimen dimensions. Thus, the upper limit takes a value of $L < 2 \cdot W/3$, and the lower limit takes a limit of $L \geq t$.

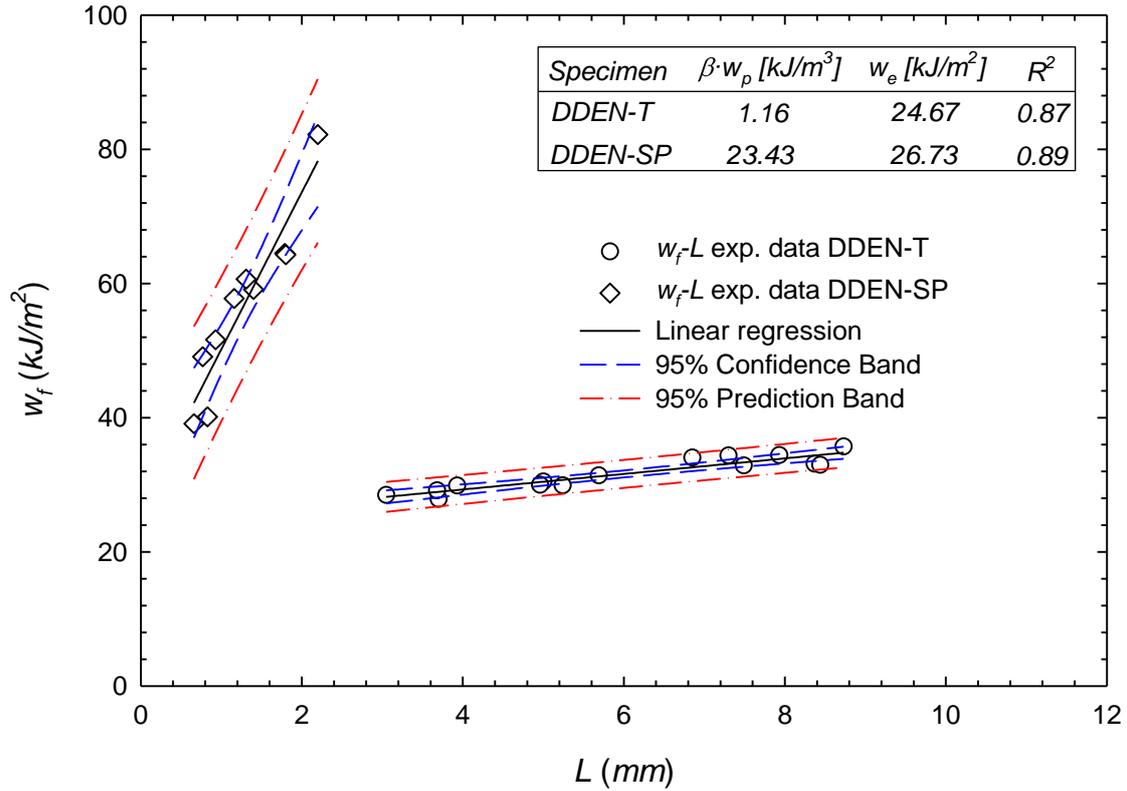

*Figure 8. Specific work of fracture, $w_f$, versus ligament length L. Comparison of the EWF results obtained for DDEN-T and DDEN-SP experiments on a fibre-reinforced polymer (Onyx).*

Regarding the analysis of the results. First, note that the specific work of fracture lines up very well in both cases. Also, note that a larger plastic work is predicted by the DDEN-SP experiment. This is intrinsically related to the test configuration, which exhibits a mainly biaxial stress state that favours ductility and plastic deformation. In addition, the punch also induces plastic deformation in the contact region. More importantly, Figure 8 reveals a very similar value of the fracture toughness, as given by $w_e$. This can be readily seen graphically, as both fitting lines intersect at the same point on the vertical axis. Thus, an initiation toughness of 24.67 kJ/m$^2$ is attained with the standardised DDEN-T experiment, while an initiation toughness of 26.73 kJ/m$^2$ is predicted by the proposed DDEN-SP test – less than 10% difference. Thus, the results suggest that the DDEN-SP experiment, in combination with the EWF method, is suitable to characterize the fracture properties of a fibre-reinforced additively manufactured polymer, with results being quantitatively comparable to those of standard tests. On the other hand, the DDEN-SP test overpredicts the plastic work measured in the DDEN-T experiment and can, therefore, only be used in a qualitative manner. Regarding the $R^2$ term, almost and identical value is obtained for DDEN-SP and DDEN-T experiments. In both cases, the value is close to 1, emphasizing the good reproducibility of the DDEN-SP experiments on Onyx. Slightly higher values are typically obtained, in both DDEN-SP and DDEN-T tests, for conventionally manufactured samples. This is in agreement with expectations, as additive manufacturing introduces more defects into the material than thermoforming and other conventional techniques. However, this appears not to be aggravated by the use of miniature specimens in the case of the fibre-reinforced polymer under consideration.

## 4. Conclusions

We use the Essential Work of Fracture (EWF) method to characterize the mechanical and fracture properties of polymeric materials that have been obtained by fused deposition modelling (FDM)-based additive manufacturing. On the one hand, we investigate the role of fibre-reinforcement on enhancing the structural integrity behaviour of additively manufactured polymers. To this end, we work with four different materials by combining standard deeply double-edge notched tensile (DDEN-T) tests and the EWF methodology. Thus, we compare the EWF parameters obtained for Onyx, a nylon-matrix, carbon-fibre reinforced crystalline polymer, with the results obtained for three well-characterized additively manufactured polymers without fibre reinforcement: polylactic acid (PLA), polypropylene (PP), and acrylonitrile butadiene styrene (ABS). On the other hand, we assess the suitability of the Small Punch Test (SPT), a mechanical test designed for testing miniature metallic specimens, to characterize the mechanical behaviour of fibre-reinforced, additively manufactured polymers. Deeply double-edge notched small punch (DDEN-SP) tests on Onyx are conducted, and compared with the results obtained from DDEN-T experiments. Our main findings are:

▪ 3D printed samples of Onyx, a new fibre-reinforced polymer, exhibit a notable enhancement in fracture toughness relative to non-reinforced polymers used widely in additive manufacturing. This finding suggests that Onyx, and other fibre-reinforced polymers, are promising candidates for additively manufactured components that demand a higher fracture resistance.

▪ DDEN-SP experiments on additively manufactured samples reveal their suitability to employ the EWF methodology. The essential requirements for the application of the method, self-similar load-displacement curves and crack propagation in mode I, are met. Additionally, upper and lower limits for the ligament length of miniature specimens have been established. These results suggest that the DDEN-SP test has the potential to enable a cost-effective assessment of the structural integrity response of polymer-based, additively manufactured components.

▪ The use of the DDEN-SP test on Onyx reveals that the EWF-based fracture toughness measurements obtained can be directly compared to those obtained with standard DDEN-T. This finding, to be examined with other polymer types, suggests that DDEN-SP fracture measurements can be directly correlated with standardised measurements, enabling a quantitative assessment.

## 5. Acknowledgements

The authors wish to thank the funding received from the Ministry of Education of the Regional Government of Castile and Leon, which started in 2018 under the support of the Recognized Research Groups of the Public Universities of Castile and Leon (project: BU033G18). E. Martínez-Pañeda additionally acknowledges financial support from Wolfson College Cambridge through their Junior Research Fellowship programme. I.I. Cuesta wishes to thank the Department of Engineering of Cambridge University for providing hospitality during his research stay.